\documentclass[11pt,twoside]{atmp}

\usepackage{amsmath,amssymb}

\usepackage[all]{xy}
\usepackage{color}

\begin{document}

\title[Symplectic Cosets]
{Bosons Live in Symplectic Coset Spaces}


\author[Eichinger]{B. E. Eichinger}

\address{Department of Chemistry \\ University of Washington \\ Seattle, WA, 98118, USA} 
\addressemail{eichinger@chem.washington.edu}

\noindent

\begin{abstract}

A theory for the transitive action of a group on the configuration space of a system of fermions is shown to lead to the conclusion that bosons can be represented by the action of cosets of the group.  By application of the principle to fundamental, indivisible fermions, the symplectic group $Sp\left( n \right)$ is shown to be the largest group of isometries of the space.  Interactions between particles are represented by the coset space $Sp\left( n \right)/ \bigotimes _1^n Sp\left( 1 \right)$.

\end{abstract}

\cutpage

\maketitle

\section{Introduction}
The observable universe is an isolated system.  This is nothing other than the statement that all observable phenomena originate within the universe and do not come from elsewhere.  To explore the implications of the notion of an isolated system, it may be useful to think about something a bit smaller than the entire universe.  Consider a system, as simple or complex as you like, embedded in an otherwise empty space (of any dimension).  Is the system translating with uniform velocity?  Clearly there is no observation that can be made from the system to determine if it is translating, as there are no reference points external to the system with respect to which an observer might verify that its position changes with time.  The notion of the position of the system in an empty space has no meaning. 

One might next ask, "Is the system rotating?" Again the answer must be 
negative, as no external reference frame exists that enables verification of this, or any other motion, of the entire system.  Certainly we can imagine the motion of the parts of the system relative to one another.  Let us suppose that there exists a group $G$ that acts naturally and transitively on the configuration space, $x$, of this system.  In acting transitively, every configuration of the system is accessible from a standard configuration by the action of an element of the group.

\section {Basic Construction}
Suppose now that one has constructed the group $G_a$ for a system $S_a$.   Using a similar procedure we can imagine constructing a group $G_b$ for $S_b$, treating it as an isolated system while ignoring the presence of $S_a$.  Now suppose the systems to be combined into $S_{ab}  = S_{a \cup b}$.  The group $G_{ab}  = G_{a \cup b}$ acts on the combined system, and in so doing encompasses all its internal motions. The product, $G_a \times G_b$, acts on the independent individual systems, each enjoying internal motions as if the other were not present.  The coset $G_{ab}/G_a  \times G_b$  therefore moves the parts of $S_a$ because of the presence of $S_b$ and \emph {vice versa}.  This obviously means that there is an interaction between the two systems, so the coset $G_{ab}/G_a  \times G_b$   represents these interactions.  Clearly this argument extends to a system consisting of several subsystems, $S_a , S_b , \cdots ,S_z $.  The motions of the parts of the 
various subsystems that result from the presence of the others will be described by the coset $G_{ab\cdots z}/G_a  \times G_b  \times  \cdots  \times G_z $. Bosons have a $G/H$ interpretation{\cite {Weinberg}, and we will generalize this concept.

This argument can be used for a system of $n$ particles.  Suppose that these are fermions, and that the properties of each are described by some function $\phi _k \left( x \right)$.  The group $G_n $ is assumed to have a natural action on the coordinates of the configuration space $x$ that is both continuous and transitive.  By the construction in the previous paragraph, we are required to define a group, $G_1$, that acts on each fermion as if it is an isolated system, so that a coset describes the interactions between the particles.  What property or properties of an individual, isolated, fermion can be defined solely within the context of a continuous (Lie) group?  The only non-trivial possibility is \emph{spin}.  On referring back to the first paragraph for discussion of isolated systems, one has to conclude that charge and mass are not intrinsic properties of isolated particles, but instead require an interaction with the surroundings (or observer) to define or quantify them.  (In the usual picture, a charge or mass generates a field, but to measure a field requires the presence of a test particle outside the system.  Inertial mass can only be defined if motion relative to a fixed frame is verifiable or that an external force is acting, and gravitational mass has meaning only where an external test mass detects its presence.  While this argument may seem to be too classical, mass and charge are classical concepts, and the language is appropriate for the concepts.)  Spin is the highly non-classical answer to the question posed above: our group $G_1$ can only be $Spin\left( 3\right) \sim SU\left ( 2\right) \sim Sp\left( 1\right)$.
  
The structure that has been built is a principal bundle (\cite{K&N}, \cite{Helgason}).  The spin degrees of freedom are contained in a $G_1$ fiber sitting over each point (particle).  The interactions live on a cross section $G/H$, with $H:= \bigotimes_1^n G_1$.  Because the spin degrees of freedom are in the fiber direction, they cannot be defined solely in terms of the coordinates in a cross section. 

Which family - orthogonal, unitary, or symplectic - should we choose for $G_1$?  To decide this we look at the parent group, $G_n$, and try for the simplest answer:  we want  either $SO\left( 3n\right) \sim Spin\left( 3n\right)$ or $SU\left ( 2n\right)$ or $Sp\left( n\right)$ to conform homogeneously to the lowest order subgroups, respectively, so as to be smoothly applicable to systems of any $n$.  The dimensions of the groups are \cite{Helgason}
\begin{align*}
\textrm{dim} \left[ SO\left (3n\right) \right] = {} &3n\left ({3n-1} \right)/2 \\
\textrm{dim} \left[ SU\left (2n\right) \right] = {} &\left ({2n+1}\right)\left ({2n-1} \right) \\
\textrm{dim} \left[ Sp\left (n\right) \right] = {} &n\left ({2n-1} \right)
\end{align*}

Now consider a system consisting of a single particle and its surroundings.  In obvious notation, the interactions between the particle and its surroundings will be described by the coset $G_n/G_1 \times G_{n-1}$.  The dimensions  of the coset spaces are \cite{Helgason}
\begin{align*}
\dim \left [SO\left (3n\right) \right/SO\left (3 \right) \times SO\left ({3n-3}\right)] = {} &9\left ({n-1}\right) \\
\dim \left \{SU\left (2n\right) \right/S\left [U\left(2 \right) \times U\left({2n-2}\right)\right]\} = {} &8\left ({n-1}\right) \\
\dim \left[Sp\left (n\right) \right/Sp\left (1 \right) \times Sp\left ({n-1}\right)] = {} &4\left ({n-1}\right)
\end{align*}
(The subgroup for the unitary case is chosen as $S\left [U\left(2 \right) \times U\left({2n-2}\right)\right]$ rather than  $SU\left(2 \right) \times SU\left({2n-2}\right)$ so as to eliminate one annoying degree of freedom.)  That is, the interaction between the subject particle and each of those in the surroundings has dimension 9, 8, or 4.  Pairwise interactions with eight or nine degrees of freedom pose a significant interpretation problem, but those with four are natural.  The only realistic choice is the symplectic group.  
We have arrived at the coset space $Sp\left(n\right)/\bigotimes_1^n Sp\left(1\right)$ that describes interactions between particles.  

The symplectic group  is also denoted by $U\left({n,\mathbb{H}}\right)$, \emph{i.e.}, it is a unitary group over the quaternion ring $\mathbb{H}$.  Another important aspect of this group is that it is both symplectic and unitary: $Sp\left(n\right)\sim Sp\left(2n,\mathbb{C}\right)\cap U\left(2n,\mathbb{C}\right)$ \cite{Simon}. It will often be convenient to shorten notation, as above, to $Sp_k:= Sp\left(k\right)$. 

It will not have escaped notice that our group is compact and has no obvious connection to relativity, nor to the Lorentz or Poincar\'e group.  However, systems and subsystems move relative to one another under the action of a compact group, and that is sufficient to realize all accessible physical states.  Connections to the Lorentz group, deSitter, and Anti-deSitter (AdS) spaces will be established later.
  
\section {Bottom up Construction and Group Action}\

The coset space $Sp\left(n\right)/\bigotimes_1^n Sp\left(1\right)$ is difficult to handle directly, so a (local) parameterization via the factorization
\begin{align*}
Sp_n/\bigotimes_1^n Sp_1 ={}& \left[Sp_n/Sp_1\times Sp_{n-1}\right]\times\left[Sp_{n-1}/Sp_1\times Sp_{n-2}\right]\times\cdots \\
\times {} &\left[Sp_2/Sp_1\times Sp_1\right]
\end{align*}
enables one to build up solutions by solving the smallest problems first.  At the bottom of these nested coset spaces is $Sp\left(2\right)/Sp\left(1\right)\times Sp\left(1\right)$.  Using standard group isomorphisms this space is alternatively represented by 
\[
Sp\left(2\right)/Sp\left(1\right)\times Sp\left(1\right)=SO\left(5\right)/SO\left(3\right)\times SO\left(3\right)=SO\left(5\right)/SO\left(4\right)=S^4
\]
\emph{i.e.}, the four-sphere (\cite{Lawson},\cite{At}). The next higher coset space is $Sp\left(3\right)/Sp\left(1\right)\times Sp\left(2\right)$, and so on.  The four-sphere will be discussed at greater length later.  

The action of the respective groups on their coset spaces is given by linear fractional transformations. To show this, let the coset space $Sp_{k+1}/Sp_1\times Sp_k$ (embedded in $Sp_n$, $n>{k+1}$) be parameterized via the elements 
\[
\exp \left[{\begin{array}{*{20}c}
 0 & 0 & 0 \\
 0 & 0 & {\xi} \\
 0 & {-\xi^*} & 0 \\
\end{array}} \right]
= \left [{\begin{array}{*{20}c}
 1 & 0 & 0 \\
 0 & \left( {1-ZZ^*}\right)^{1/2} & Z \\
 0 & {-Z^*} & \left( {1-Z^*Z}\right)^{1/2} 
\end{array}} \right]
\]
derived from the Lie algebra \cite{Gilmore}.  Here $\xi$ is a $k$-dimensional vector over the quaternions, and $Z = \left(\xi\xi^*\right)^{-1/2}\left[\sin\left(\xi\xi^*\right)^{1/2}\right]\xi = \xi \left(\xi^*\xi\right)^{-1/2}\sin\left(\xi^*\xi\right)^{1/2}$.   The conjugate transpose of a quaternion vector (or matrix) $x$ is denoted by $x^*$.  [For ${k+1}<n$ this representation is embedded in the $n\times n$ larger matrix as shown; in the sequel this embedding will be understood and the padding will be omitted.]  

An element $g\in Sp_{k+1}$ expressed in a conforming partitioning is  
\[
g=\left[{\begin{array}{*{20}c}
 A & B \\
 C & D
 \end{array}}\right] ;  \quad g^{-1} = g^* = \left[{\begin{array}{*{20}c}
  A^* & C^* \\
  B^* & D^*
  \end{array}}\right],
 \]
and this acts by
\[
gxH=yH
\]
\[
gxH=\left[{\begin{array}{*{20}c}
 A & B \\
 C & D
 \end{array}}\right]
 \left [{\begin{array}{*{20}c}
 \left( {1-ZZ^*}\right)^{1/2} & Z \\
 {-Z^*} & \left( {1-Z^*Z}\right)^{1/2} 
\end{array}} \right]H
\]
\[
yH=\left [{\begin{array}{*{20}c}
 {A\left( {1-ZZ^*}\right)^{1/2}-BZ^*} & {AZ+B\left(1-Z^*Z\right)^{1/2}} \\
 {C\left( {1-ZZ^*}\right)^{1/2}-DZ^*} & {CZ+D\left( {1-Z^*Z}\right)^{1/2}} 
\end{array}} \right]H
\]
where $H\sim Sp_1\times Sp_k$.  This construction works for a larger class of coset spaces than is implied here; $Z$ might just as well belong to $Sp\left( k\right)/Sp\left( j\right)\times Sp\left({k-j}\right)$.  At several places ahead this larger coset space will be intended in the development.  

Physically what is happening is that the system, the $Sp\left(j\right)$ part, and the surroundings, the $Sp\left({k-j}\right)$ part, may experience an arbitrary change in their spin/internal states as a result of the action of $g$, and this change is conveyed by the subgroup $H$.  We want to know how $g$ acts on the coset space, and this requires that the action of $H$ be eliminated by taking ratios.  The action of $G\sim Sp\left( k\right) $ on $G/H\sim Sp\left( k\right) /Sp\left( j\right)\times Sp\left({k-j}\right)$, $\left({g: Q\to P}\right)$,  is more succinctly represented by
\begin{equation} \label {lft} 
P =  \left({AQ+B}\right)\left({CQ+D}\right)^{-1} \\
 = \left({-QB^*+A^*}\right)^{-1}\left({QD^*-C^*}\right)\
 \end{equation}
with $Q=Z\left({1-Z^*Z}\right)^{-1/2}$.  Given this action it is seen that the origin of the inhomogeneous space  is mapped to $g: O\to BD^{-1}=-\left(A^*\right)^{-1}C^*$, which is useful for some computations.  Clearly the coset space has the appearance of being non-compact, as a point satisfying ${CQ+D}=0$ or $\left({-QB^*+A^*}\right)=0$ is mapped to infinity.  

\section{Metric and Curvature}

This section is standard material and only the results will be presented \cite{Hua}.  The invariant line element on $Sp\left( k\right) /Sp\left( j\right)\times Sp\left({k-j}\right)$ is given by 
\begin{align}\label {metric}
ds^2 = {} & \textrm{tr}\left[\left({1+QQ^*}\right)^{-1}dQ\left({1+Q^*Q}\right)^{-1}dQ^*\right] \\
  = {} & \textrm{tr}\left[\left({1+QQ^*}\right)^{-1}dQdQ^*-\left({1+QQ^*}\right)^{-1}dQQ^*\left({1+QQ^*}\right)^{-1}QdQ^*\right] 
\end{align}
where the second version follows from $\left({1+Q^*Q}\right)^{-1}=1-Q^*\left({1+QQ^*}\right)^{-1}Q$.  Calculation of the Maurer-Cartan form (\cite{K&N}, \cite{Chern}) for the general case, $Sp_k/Sp_j\times Sp_{k-j}$,  gives the curvature form as 
\begin{equation} \label{curve}
\Omega = \left[{\begin{array}{cc}
{h_j^*} & 0 \\
0 & {h_{k-j}^*}
\end{array}} \right]
\left[{\begin{array}{cc}
{d\mathcal Q\wedge d\mathcal Q^*} & 0 \\
0 & {d\mathcal Q^* \wedge d\mathcal Q} 
\end{array}} \right]
\left[{\begin{array}{cc}
h_j & 0 \\
0 & h_{k-j} 
\end {array}}\right].
\end{equation}
Here $d\mathcal Q = \left({1+QQ^*}\right)^{-1/2}dQ\left({1+Q^*Q}\right)^{-1/2}$, $h_j\in Sp_j$ and $h_{k-j}\in Sp_{k-j}$.  This identifies the $Sp_k/Sp_j\times Sp_{k-j}$ coset space as an Einstein space.  The form encompasses both self-dual and anti-self-dual sectors for $n = 2$ \cite{At}.

\section{Equation of Motion and Lie Algebra}

The development to this point has been vague as to the object that describes the particles that the group is acting upon.  This will now be rectified.  Let 
\[
\Phi\left(xH\right)=\left[{\begin{array}{cc}
\phi_a\left(xH\right) \\
\phi_b\left(xH\right)
\end{array}}\right]
\]
be a square-integrable vector-valued function that is compatible with 
the matrix representation of the group that we have been working with.  This is a vector bundle associated to the principle bundle (\cite{K&N},\cite{Darling}).  One might think of $\phi_a\left(xH\right)$ as the wave function of the system, and $\phi_b\left(xH\right)$ as that of the surroundings.  The most natural equation of motion that is consistent with quantum theory and our group action is provided by a one-parameter local group action with Lie algebra $\mathfrak{g}$, so that
\begin{equation} \label {motion}
\partial \Phi/\partial t = \mathfrak{g}\Phi
\end{equation}
where $\mathfrak{g}$ consists of the infinitesmal generators of the Lie algebra.  The parameter $t$ is safely identified with Galilean 
time.  [We are using natural units with $\hbar=c=1$.  Furthermore, skew-symmetry of the Lie algebra, with concomitant suppression of $\sqrt{-1}$, is preferred over the Hermitean option so as to keep equations simple.]  

We now need to fix a representation for the quaternions so as to present the Lie algebra.  Let the basis for the quaternions consist of $\{\mathbf{e,i,j,k}\}$, with 
\[
\mathbf{e}^2=\mathbf{e}=1; \quad \mathbf{i}^2=\mathbf{j}^2=\mathbf{k}^2=-1; \quad \mathbf{ij}=\mathbf{k}, \quad \mathbf{jk}=\mathbf{i},\quad \mathbf{ki}=\mathbf{j}
\]
A quaternion $q$ will be written $q:=w\mathbf{e}+x\mathbf{i}+y\mathbf{j}+z\mathbf{k}$, with conjugate $q^*:=w\mathbf{e}-x\mathbf{i}-y\mathbf{j}-z\mathbf{k}$.  The norm $|q|$  of $q$ is defined by $|q|^2=qq^*=q^*q=w^2+x^2+y^2+z^2$.  The derivative is most naturally defined such that $dq/dq = 1$, which implies that 
\[
d/dq=\frac{1}{4}\left(\mathbf{e}\partial/\partial{w}- \mathbf{i}\partial/\partial{x}-\mathbf{j}\partial/\partial{y}-\mathbf{k}\partial/\partial{z}\right).
\]
The representation of a quaternion in a basis of Pauli matrices (modulo $i=\sqrt{-1}$ ) is 
\[
q = \left[{\begin{array}{cc}
{w+iz} & {x + iy} \\
{-\left(x-iy\right)} & {w-iz}
\end{array}}\right]=\left[{\begin{array}{cc}
{\rho_1} & {\rho_2} \\
{-\bar\rho_2} & {\bar\rho_1}
\end{array}}\right]
\]
in this representation 
\begin{equation}\label{quat}
q^*=\bar{q}^{\prime}=\left[{\begin{array}{cc}
{\bar\rho_1} & {\bar\rho_2} \\
{-\rho_2} & {\rho_1}
\end{array}}\right]^{\prime}=\left[{\begin{array}{cc}
{\bar\rho_1} & {-\rho_2} \\
{\bar\rho_2} & {\rho_1}
\end{array}}\right]
\end{equation}
where $\bar{r}$ is the complex conjugate and $r^{\prime}$ is the transpose of the matrix $r$ . The differential operator in the matrix representation is 
\begin{equation}\label{deriv}
d/dq=\frac{1}{2}\left[{\begin{array}{cc}
{\partial/\partial{\rho_1}} & {-\partial/\partial{\bar\rho_2}} \\
{\partial/\partial{\rho_2}} & {\partial/\partial{\bar\rho_1}}
\end{array}}\right]
\end{equation}
There does not seem to be a choice for the representation of $d/dq$ that simultaneously satisfies $dq/dq = 1$ and $dq^*/dq =0$.  For this choice of  $d/dq$, eq. (\ref{deriv}), one finds $dq^*/dq = -1/2$.  

An important automorphism is provided by $J_1 = \mathbf{i}$ acting by conjugation $J_1\cdot q \to J_1^{-1}qJ_1 = \bar q$.  Define $J_k:=\mathbf{1}_k\bigotimes J_1$ where $\mathbf{1}_k$ is the identity matrix of rank $k$.  Conjugation of a $k$-dimensional quaternion-valued vector $Q$ is provided by $\bar Q = J_1^{-1}QJ_k$, so that $ Q^* = J_kQ^{\prime}J_1^{-1}$. 
	
The components of the Lie algebra, $\mathfrak{g}\in sp_{k+1}$, that are parameterized by the coordinates of $Sp_{k+1}/Sp_1\times Sp_k$ have infinitesmal generators given by 
\begin{align*}
\mathfrak{h}_1= {} & Q\partial{}^{\prime}-\left(Q\partial{}^{\prime}\right)^{\prime} \\
\mathfrak{h}_k =  {} & Q^*\bar\partial{}-\left(Q^*\bar\partial{}\right)^* \\
\mathfrak{p} =  {} & \bar\partial{}+ Q\left(Q^*\bar\partial{}\right)^* \\
  =  {} & \left(1+QQ^*\right)\bar\partial{}+Q\mathfrak{h}_k \\
\end{align*}
Here $\partial{}=\left(\partial{}/\partial{q_1}, \partial{}/\partial{q_2},\cdots, \partial{}/\partial{q_k}\right)$.  A tedious but straightforward calculation shows that these generators follow the canonical commutation relations \cite{Helgason}
\[
\left[\mathfrak{h}_a,\mathfrak{h}_a\right]\in \mathfrak{h}_a;\quad \left[\mathfrak{h}_a,\mathfrak{h}_b\right]=0; \quad \left[\mathfrak{h}_a,\mathfrak{p}\right]\in \mathfrak{p}; \quad \left[\mathfrak{p},\mathfrak{p}\right]\in \{\mathfrak{h}_a,\mathfrak{h}_b \}.
\]
Here $\{\mathfrak{h}_a,\mathfrak{h}_b \}$ is the set of operators in the subgroup - not the anti-commutator.  The Laplace-Beltrami operator associated to these generators, $ \Delta$, is the trace of the square of the matrix of generators.  Thus 
\[
\Delta = \textrm{tr}\left(\mathfrak{h}_1\mathfrak{h}_1^*+\mathfrak{p}\mathfrak{p}^*\right)+\textrm{tr}\left(\mathfrak{h}_k\mathfrak{h}_k^*+\mathfrak{p}^*\mathfrak{p}\right).
\] 
This is only one among many operators that can be formed from the generators of the Lie algebra.

On returning to eq. (\ref {motion}) one notes that the eigenvalues of a symplectic matrix are pure imaginary (\cite{Simon},\cite{Fulton}), which recommends that we look for eigenvector solutions of the equation with $\Phi\left(t,xH\right) = \Phi\left(xH\right)exp\left(iEt\right)$.  The equation to be solved is thus 
\begin{equation} \label {eig}
iE\left[{\begin{array}{cc}
\phi_1\left(xH\right) \\
\phi_k\left(xH\right)
\end{array}}\right]=
\left[{\begin{array}{cc}
   {\mathfrak{h}_1} &{-\mathfrak{p}} \\
 {\mathfrak{p}^*} & {\mathfrak{h}_k}
\end{array}}\right]\left[{\begin{array}{cc}
\phi_1\left(xH\right) \\
\phi_k\left(xH\right)
\end{array}}\right].
\end{equation}
Furthermore, because the eigenvalues of a representation of the symplectic group occur in conjugate pairs \cite{Simon}, both positive and negative energy solutions will be obtained from eq. (\ref{eig}).  This may have implications for the interconversion of matter and anti-matter.  A plane wave solution of eq. (\ref{eig}) at the origin ($Q=0$) gives the relativistic energy of the particle, provided the identity component of the quaternion is conjugate to rest mass and the imaginary components are conjugate to momenta.

There is a further remarkable feature of eq. (\ref{motion}) perhaps best described in the time dependent form (with $t$ implicit in functions).  Writing out the right hand side one has  
\begin{equation}\label {}
\left[{\begin{array}{cc}
\partial\phi_1\left(xH\right)/\partial t \\
\partial\phi_k\left(xH\right)/\partial t
\end{array}}\right]=
\left[{\begin{array}{cc}
{\mathfrak{h}_1}\phi_1\left(xH\right) -\mathfrak{p}\phi_k\left(xH\right) \\
{\mathfrak{p}^*}\phi_1\left(xH\right)+ \mathfrak{h}_k\phi_k\left(xH\right)
\end{array}}\right].
\end{equation}
At an instant of observation, as registered by the change in the wave function $\partial\phi_k\left(xH\right)/\partial t$ for the surroundings, the system $\phi_1\left(xH\right)$ reports its \emph {present} state via the operator $\mathfrak{p}^*$.  However, since the system is in contact with its surroundings \emph {via} the $\mathfrak{p}\phi_k\left(xH\right)$ term, its state will evolve and the next observation will find the system in a different state.  [While the equation is written for a single particle, the argument works for many-body systems as well.] Whether the change of state is large or small depends on the strength of the coupling, and that is a problem for another time.

The $\mathfrak {h}_1$ operator in eq. (\ref{eig}) couples the spin of the particle to the surroundings.  In the fiber bundle picture what is happening is that the Lie algebra transports the system in a tangent plane, and this projects onto a motion along the $Sp\left(1\right)$ fiber (since a cross section of the bundle is not flat).  Were it not for this coupling, particle spin could not be observed or manipulated. [The difficulty of combining into words or pictures the spin of an electron with its point-like extension is a direct consequence of this fiber bundle structure.]

\section{The Alternative Representation of $Sp\left(n\right)$}

To this point the theory has been developed in the representation $U\left(n,\mathbb{H}\right)$.  Given the fact that quaternions are not commutative and have had a bad name since the time of Gibbs, some may find it more comfortable to work in the $Sp\left(2n,\mathbb{C}\right)$ version.  The conjugation operation $q^*=J_1^{-1}q^{\prime}J_1$ provides just what is needed to map between the two representations\cite{Varad}.  For $g\in U\left(n,\mathbb{H}\right)$ we have $g^*g = J_n^{-1}g^{\prime}J_n g = 1$, so that $g^{\prime}J_n g = J_n$, where $J_n:=\mathbf{1}_n\bigotimes J_1$ as before.  Now, there exists a permutation $\mathcal{P}$ such that $\mathcal{P}: \left(\mathbf{1}_n\bigotimes J_1\right) \to J_1\bigotimes \mathbf{1}_n$, and acting on $g$ gives a permuted form $\mathcal{P}: g \to \mathcal{G}$.  For $\mathcal{J}=J_1\bigotimes \mathbf{1}_n$, this gives $\mathcal{G^{\prime}JG}=\mathcal{J}$, which is the standard definition of the symplectic group over the complex numbers [$\mathcal{G} \in Sp\left(2n,\mathbb{C}\right)$]; the group preserves a skew-symmetric bilinear complex form.  But the group is also unitary, \emph{i.e.}, $Sp\left(n\right)\sim Sp\left(2n,\mathbb{C}\right)\cap U\left(2n,\mathbb{C}\right)$, as noted above, so that we also have $\mathcal{G^*G}=1$.  These two properties yield the Lie algebra in this representation in the form
\[
\mathfrak{g}\sim \left[{\begin{array}{cc}
\mathfrak{a} & \mathfrak{b} \\
{-\mathfrak{b}^*} & {-\mathfrak{a}^{\prime}}
\end{array}}\right]; \quad \mathfrak{a}^* = -\mathfrak{a}; \quad \mathfrak{b}^{\prime} = \mathfrak{b}
\]
where $\mathfrak{a}$ and $\mathfrak{b}$ are complex matrices.  Cosets in this representation are messier to work with than those in the quaternion 
basis, so this representation is not pursued further here.  However, one should note that the $\mathfrak{a}$-sector of this representation, taken alone, 
is isomorphic to $U\left(n\right)$.  

\section {The Action of $SL\left(2,\mathbb{C}\right)/\{\pm\mathbf {1}\}\sim SO\left(3,1\right) $}

A thorough study of the relation between symplectic cosets and traditional field theories will take considerable effort.  As a start, one simply notes that in generally accepted field theories all of the interactions between a particle and its surroundings take place in a four-dimensional space-time manifold rather than in this pairwise additive $4n$-dimensional setting.  Why hasn't the symplectic group asserted itself?  Perhaps the answer is that $u\in SL\left(2,\mathbb{C}\right)$ acting by conjugation on any quaternion $q$ by $u: q \to u^{-1}qu$ preserves norm and trace, and therefore $SL\left(2,\mathbb{C}\right)$ can act independently on each and every element in a row/column of $g\in\ Sp\left(n\right)$ and not disturb normalization.  [However, this action will, in general, perturb orthogonality.  But an anomaly that might become evident as a result of broken orthogonality would only show up in three body interactions.  In the absence of a complete solution of the three body problem, this could easily have gone undetected.]  Since this action of $SL\left(2,\mathbb{C}\right)$ by conjugation lifts to the algebra $sp\left(n\right)$, and preserves the trace, the identity component of the quaternion (in the algebra) once again appears to be conjugate to rest mass.  [For this discussion the action of $SL\left(2,\mathbb{C}\right)$ might also have been of the norm preserving form $aqb$ for $\{a,b\}\in SL\left(2,\mathbb{C}\right)$.  But then another interpretation for the identity component of the quaternion would have to be found.]

\section {Averaging Interactions: The Large N Limit}

As mentioned earlier, contact with field theories will at least require that the  metric associated to the $n$-dimensional quaternion $Q$ be simplified to one quaternion dimension.  The most straightforward way to do this is by averaging. For the  
$Sp_{n+1}/Sp_1\times Sp_n$ metric in eq. (\ref{metric}) one may introduce the averages 
\begin{equation}\label{ave}
n<d\sigma d\sigma^*>=dQdQ^*; \quad n<\sigma \sigma^*>=n|\sigma|^2 = QQ^*; \quad n<\sigma d\sigma^*> = QdQ^*
\end{equation}
so that the  metric can be written as
\begin{equation}\label{avemet}
ds^2= \textrm{tr}\{f\left(n|\sigma|^2\right)|\sigma|^{-2}\left[<d\sigma d\sigma^*>
- <\sigma d\sigma^*>^*f\left(n|\sigma|^2\right)|\sigma|^{-2}<\sigma d\sigma^*>\right]\}
\end{equation}\begin{align*}
\approx \textrm{tr}\{ \left(1-n^{-1}|\sigma|^{-2}\right)|\sigma|^{-2}\left[<d\sigma d\sigma^*>-<\sigma d\sigma^*>^*|\sigma|^{-2}<\sigma d\sigma^*>\right]\} \\
+ \textrm{tr}\{ \left(1-n^{-1}|\sigma|^{-2}\right)|\sigma|^{-2}<\sigma d\sigma^*>^*n^{-1}|\sigma|^{-4}<\sigma d\sigma^*>\}.
\end{align*}
Here $f\left(x\right)=\left(1+1/x\right)^{-1}$; $\sigma$ and $d\sigma$ are understood to be one-dimensional (resp. infinitesmal) quaternions (four real dimensions) that are defined in eq. (\ref{ave}).  The second, approximate, version of the metric is valid for large $n$ only.  The metric on $Sp_2/Sp_1\times Sp_1$ that we are trying to emulate is 
\begin{equation}\label{s4}
ds^2=\textrm{tr}\left[\left(1+|q|^2\right)^{-2}dqdq^*\right].
\end{equation}
(Note that this is the metric for an instanton (\cite{Lawson},\cite{At}). The first term in the approximate version of eq. (\ref {avemet}) can be made to vanish (at least at a point) by choosing the $\rho = \sigma/|\sigma |$ part of $<\sigma d\sigma^*>$ such that $<\rho d\sigma^*><\rho d\sigma^*>^* \approx <d\sigma d\sigma^*>$.  To first order in $n^{-1}$ the metric then reduces to
\[
 ds^2\approx n^{-1}\textrm{tr}\left[|\sigma|^{-2}<\sigma d\sigma^*>]\sigma|^{-4}<\sigma d\sigma^*>^*\right] \approx n^{-1}\textrm{tr}\left[|\sigma|^{-4}d\sigma d\sigma^*\right],
 \]
which seems to be as close as one can get to eq. (\ref{s4}) (corresponding to q large in that equation).  This bears some resemblance to the large N limit of metrics considered by Maldacena \cite {Mal}.  More careful approximations for small systems embedded in large N surroundings will be very important to understanding how this theory works. 

\section {$S^4$ and the Conformal Group} 

The isomorphism $Sp\left(2\right)/Sp\left(1\right)\times Sp\left(1\right) \sim SO\left(5\right)/SO\left(4\right) \sim S^4$ enables one to use real coordinates for calculations: the geometry is, of course, just 
\[
\sum\limits_{i=0}^4{x_i^2}=1.
\]
$SO\left(5\right)$ acts by linear fractional 
transformations on the inhomogeneous coordinates $y_k=x_k/x_0, 1\le k \le 4$, with $1+yy^{\prime}=1/x_0^2 \ge 1$.  Evaluation of the invariant metric is a standard calculation, yielding the line element in the $y$-coordinates as 
\[
ds^2 = \left(1+yy^{\prime}\right)^{-1}dy\left(1+y^{\prime}y\right)^{-1}dy^{\prime}.
\]
The relation between the metrics on $Sp\left(2\right)/Sp\left(1\right)\times Sp\left(1\right)$ and $SO\left(5\right)/SO\left(4\right)$ is not trivial, and is left for another time.  In any case, one may just as well use the standard spherical line element on the 4-sphere for calculations.  The self-dual and anti-self-dual connections, Yang-Mills action, and instantons are more accessible from the quaternion version (\cite{Lawson},\cite{At}).
  
\subsection {What's Inside/Outside the Sphere?}

The 4-sphere is the surface of the 5-ball, $B^5$.  Since our larger group ($Sp_n$) action admits inversions, one is encouraged to look also at the exterior of the sphere, obtained by inverting the space through the $S^4$ surface.  The extended Lorentz group $SO\left(1,5\right)$ acts on the ball 
(a de Sitter space), while the anti-de Sitter (AdS) space ($vv^{\prime}-v_0^2 = 1 \Rightarrow xx^{\prime}-1>0;x=v/v_0$) is the inverted ball (here $v$ is a 5-dimensional vector).   The group $SO\left(1,5\right)$ is the conformal group of $S^4$ (\cite {Mal},\cite{Wit}).  The boundary, $S^4$, of either $B^5$ or its inverse is approached in the limit $v_0 \to \infty$. No conjecture is offered as to what lives in the de Sitter or AdS space. 

\section {Representations}

The action of the symplectic group through its representations has  an interesting implication.  Select a $g \in Sp\left(n\right)$  and a 
representation $A_g$ acting on functions $\Phi\left(xH\right)$ according to $A_g \Phi\left(xH\right) = \Phi \left(g^{-1}xH\right)$.  This equation asserts is that we can know the current state of the system (modulo H) only by knowing the past.  The differential form, eq. (\ref{motion}), enables one to project a short time into the future, but no more than that.    It is probable that induced representation theory ({\cite{Simon},\cite {Folland}) will be useful in climbing up the nested coset spaces.
 
The real work in this theory will be done with higher dimensional representations that have the capacity to encompass the exchange of excitations between systems (and surroundings).  It might be useful to begin with the conjecture that the dimension of the representation of $Sp_m$ that acts on an isolated system is fixed, but that excitations can be passed 
between $Sp_k\times Sp_{m-k}$ subsystems.    No further conjecture is offered as to the mechanism for doing this.  While exercising restraint in making additional conjectures, there is one more that is too tantalizing to pass up.  This is the possibility that the "bubbling-off theorem" of four-manifolds may offer a topological picture of particle production (\cite{Lawson},\cite{Mal2}).  

\section {Conclusions}
	
This has been a condensed presentation of a comprehensive theory of interactions - it is only a beginning.  Correspondences with geometries of current interest in particle theory have been discussed where appropriate, and a definite fiber bundle structure has been established to treat many-body interactions.  This formulation of a group action has the potential to provide new insights into the structure of matter.  First and foremost amongst these is the notion that to infer the properties of the simplest particle beyond its intrinsic angular momentum it is necessary to couple the particle to its surroundings, not to the vacuum.  Symplectic coset spaces are necessary to understand many-body quantum interactions.   Only by solving the equations will one learn if they are sufficient.

\bigskip
\textsc{Acknowledgment}

The author is grateful to Ulrich Suter and Peter G\"{u}nter of the ETH for their hospitality during the Spring of 2008.  A portion of this work was accomplished at that time.  Gerald Folland and John Sullivan answered several mathematical questions.

\bibliographystyle{my-h-elsevier}

\end{document}